\documentclass[runningheads]{llncs}
\usepackage{graphicx}
\usepackage{amsmath,amssymb,mathrsfs,framed,esint,slashed}
\usepackage[colorlinks]{hyperref}
\usepackage{color}
\usepackage[all]{xy}

\newcommand{\de}{{\rm d}}

\begin{document}
\title{From quantum hydrodynamics\\to Koopman wavefunctions I\thanks{CT acknowledges partial support by the Royal Society and the Institute of Mathematics and its Applications, UK.}
}
%
%
\author{
\makebox{Fran\c{c}ois Gay-Balmaz\inst{1}
\and Cesare Tronci\inst{2}
}}
\authorrunning{F. Gay-Balmaz and C. Tronci}
%
\institute{$^1$CNRS \& \'Ecole Normale Sup\'erieure, Paris, France \\
\email{francois.gay-balmaz@lmd.ens.fr}
\\
\medskip
$^2$Department of Mathematics, University of Surrey, Guildford, UK\\
\makebox{Department of Physics and Engineering Physics, Tulane University, New Orleans, USA}\\
 \email{c.tronci@surrey.ac.uk}
}
\maketitle              

\begin{abstract} Based on Koopman's theory of classical wavefunctions in phase space, we present the Koopman-van Hove (KvH) formulation of classical mechanics as well as some of its properties. In particular, 
we show how the associated classical Liouville density arises as a momentum map associated to the unitary action of strict contact transformations on classical wavefunctions. Upon applying the Madelung transform from quantum hydrodynamics in the new context, we show how the Koopman wavefunction picture is insufficient to reproduce arbitrary classical distributions. However, this problem is entirely overcome by resorting to von Neumann operators. Indeed, we show that the latter  also allow for singular  $\delta-$like profiles of the Liouville density, thereby reproducing point particles in phase space.

\end{abstract}

\section{Introduction}
Koopman wavefunctions on phase-space have a long history going back to Koopman's early work  \cite{Koopman} and their unitary evolution was revisited by van Hove  \cite{VanHove}, who unfolded their role within a niche area of symplectic geometry now known as \emph{prequantization} \cite{Kirillov,Kostant,Souriau}. In this context, classical mechanics possesses a Hilbert-space formulation similar to quantum mechanics and our recent work \cite{BoGBTr} provided a new geometric insight on the correspondence between Koopman wavefunctions and classical distributions on phase-space. Then, classical wavefunctions were renamed \emph{Koopman-van Hove}  wavefunctions to stress van Hove's contribution.

Here, we show how the evolution of Koopman wavefunctions can be described in terms of the same geometric quantities already emerging in Madelung's hydrodynamic formulation of standard quantum mechanics \cite{Madelung}. In particular, after reviewing the geometry of quantum hydrodynamics, we shall show how the introduction of density and momentum variables on phase-space leads to an alternative geometric description of Koopman classical mechanics. While most of our discussion is based on \cite{GBTr} (where further details can be found), here we shall also show that this picture naturally allows for $\delta-$like  particle solutions, whose occurrence is instead questionable in the Hilbert-space wavefunction picture.

\section{Geometry of quantum hydrodynamics}\label{section_1}

Madelung's equation of quantum hydrodynamics are obtained by replacing the polar form $\psi(t,x)=\sqrt{D(t,x)}e^{-{\rm i}S(t,x)/\hbar}$ of the wavefunction into Schr\"odinger's equation ${\rm i}\hbar\partial_t\psi=-m^{-1}\hbar^{2}\Delta\psi/2+V\psi$. 
Then, defining the velocity vector field $v= \nabla S/m$ leads to the well-known set of hydrodynamic equations
\begin{equation}\label{MadelungEqs2}
\partial _t v+ {v}\cdot\nabla v=-\frac1m\nabla\bigg(  V +\frac{\hbar^2}{2m}   \frac{\Delta  \sqrt{D}}{ \sqrt{D}}\bigg) \qquad  \quad 
\partial _t D+\operatorname{div}( Dv)= 0\,.
\end{equation} 
Madelung's equations were the point of departure for Bohm's interpretation of quantum dynamics \cite{Bohm}, in which the integral curves of 
$v(x,t)$ are viewed as the genuine trajectories in space of the physical quantum particles. The fact that a hydrodynamic system arises from the Schr\"odinger equation is actually a consequence of an intrinsic geometric structure underlying the hydrodynamic density and momentum variables $D$ and $\mu=D\nabla S$.

\smallskip\noindent
{\bf Madelung's momentum map.} Madelung's  equations \eqref{MadelungEqs2} can be geometrically explained by noticing that the map
\begin{align}\nonumber
&\mathcal{J} :L^2(M, \mathbb{C} ) \rightarrow \mathfrak{X} (M)^* \times \operatorname{Den} (M)\,, \quad 
\\
&\mathcal{J} ( \psi )=\big(\hbar\operatorname{Im}(\psi^*\nabla\psi),|\psi|^2\big)=(mDv,D)
\label{Madmomap}
\end{align}
is an equivariant momentum map. Here,  $L^2(M, \mathbb{C} )$ is the Hilbert space of quantum wavefunctions, while $\mathfrak{X} (M)^* \times \operatorname{Den} (M)$ is the dual space to the semidirect-product Lie algebra $\mathfrak{X}(M)\,\circledS\,\mathcal{F}(M)$. The momentum map structure of \eqref{Madmomap} is associated to the following unitary representation of the semidirect-product group $ \operatorname{Diff}(M) \,\circledS\, \mathcal{F} (M, S ^1 )$ on $L^2(M, \mathbb{C} )$:
\begin{equation}\label{QuantumMadelungRep}
\psi\mapsto \sqrt{J_\chi^{-1}}\,({e^{-{\rm i}\varphi/\hbar}}\psi) \circ \chi ^{-1} , \qquad ( \chi  , e^{{\rm i} \varphi} ) \in \operatorname{Diff}(M) \,\circledS\, \mathcal{F} (M, S ^1 )
\,,
\end{equation} 
where $J_\chi=\det\nabla \chi $ is the Jacobian determinant of $\chi$.
The notation is as follows: $\operatorname{Diff}(M)$ is the group of diffeomorphisms of $M$, $\mathfrak{X}(M)$ is its Lie algebra of vector fields, and $\mathcal{F}(M, S ^1 )$ is the space of $S^1$-valued functions on $M$.
Being an equivariant momentum map, $ \mathcal{J} $ is a Poisson map hence mapping Schr\"odinger's Hamiltonian dynamics on $L^2(M, \mathbb{C})$ to Lie-Poisson fluid dynamics on  $\mathfrak{X} (M)^* \times \operatorname{Den} (M)$.

\smallskip\noindent
{\bf Definition of momentum maps.} Without entering the technicalities involved in the construction of semidirect-product groups, here we simply recall the definition of momentum maps.  Given a Poisson manifold $(P, \{,\} )$ and a Hamiltonian action of a Lie group $G$ on $P$, we call $ \mathcal{J} :P \rightarrow \mathfrak{g} ^* $ a \textit{momentum map} if it satisfies 
$\{f,\langle \mathcal{J} , \xi \rangle\}=\xi _P[f]$.
Here, $\mathfrak{g}$ denotes the Lie algebra of $G$ and $\mathfrak{g}^*$ its dual space, $\xi_P$ is the infinitesimal generator of the $G-$action $\Phi_g:P\to P$ and $\langle\cdot,\cdot\rangle$ is the natural duality pairing on $\mathfrak{g}^*\times\mathfrak{g}$. A momentum map is equivariant if $\operatorname{Ad}^*_g{\mathcal{J}}(p)={\mathcal{J}}(\Phi_g(p))$ for all $g\in G$.
When $P$ is a  symplectic vector space carrying a (symplectic) $G-$representation with respect to the  symplectic form $\Omega$, the momentum map $\mathcal{J}(p)$ is given by $2\langle \mathcal{J}(p),\xi\rangle:=\Omega(\xi_{P}(p),p)$,  for all $p\in P$ and  all $\xi\in \mathfrak{g}$. Here we specialize this definition to the symplectic Hilbert space  $P=L^2(M,\Bbb{C})$ of wavefunctions, endowed with the standard symplectic form $\Omega(\psi_1,\psi_2)=2\hbar\,{\rm Im}\int_M\bar{\psi}_1 \psi_2 \,\de x$. Then, a momentum map associated to a unitary $G-$representation  on $L^2(M,\Bbb{C})$  is the map ${\mathcal{J}}(\psi)\in\mathfrak{g}^*$  given by 
\begin{equation}\label{momap_formula}
\left<{\mathcal{J}}(\psi),\xi\right>=-\hbar\,{\rm Im}\int_{M\!}\bar\psi\,{\xi_P(\psi)}\,\de x\,.
\end{equation}  
It is easily checked that, if $G=  \operatorname{Diff}(M) \,\circledS\, \mathcal{F} (M,S ^1 )$ acts symplectically on $P= L^2(M, \mathbb{C} )$ via \eqref{QuantumMadelungRep}, expression \eqref{momap_formula} yields the Madelung momentum map.

\section{Koopman-van Hove formulation of classical mechanics}
As anticipated in the introduction, we want to show how the geometric setting of quantum hydrodynamics transfers to the case of classical wavefunctions in their Koopman-van Hove (KvH) formulation. The latter possesses a deep geometric setting which is reviewed in the present section.

\subsection{Evolution of Koopman wavefunctions\label{sec:KvH}}
Let $Q$ be the configuration manifold of the classical mechanical system and $T^*Q$ its phase space, given by the  cotangent bundle of $Q$. We assume that the manifold $Q$ is connected. The phase space is canonically endowed with the one-form ${\mathcal{A}}= p_i{\rm d}q^i$ and the symplectic form $\omega=-{\rm d}{\mathcal{A}}= {\rm d} q^i\wedge {\rm d}p_i$. For later purpose, it is also convenient to consider the trivial  circle bundle 
\begin{equation}\label{prequantumbundle}
T^*Q\times S^1\rightarrow T^*Q\,,
\end{equation}
known as \emph{prequantum bundle}. Then,  ${\mathcal{A}}$ identifies a principal connection ${\mathcal{A}}+ {\rm d} s$ with curvature given by (minus) the symplectic form $\omega$.
A classical wavefunction $\Psi$ is an element of the classical Hilbert space 
$\mathscr{H}_{\scriptscriptstyle C}=L^2(T^*Q, \mathbb{C})$ with standard Hermitian inner product
$
\langle \Psi_1| \Psi_2\rangle:= \int_{T^*Q\!} \bar{\Psi}_1\Psi_2\,{\rm d} z,
$
where ${\rm d}z$ denotes the Liouville volume form on $T^*Q$.
The corresponding real-valued pairing and symplectic form on $\mathscr{H}_{\scriptscriptstyle C}$ are defined by
\begin{equation}\label{inner_symplectic}
\langle \Psi_1, \Psi_2\rangle= \operatorname{Re} \langle \Psi_1| \Psi_2\rangle
\quad\text{and}\quad\Omega(\Psi_1, \Psi_2)= 2\hbar \operatorname{Im}\langle \Psi_1| \Psi_2\rangle
\,.
\end{equation}

\smallskip\noindent
{\bf The KvH equation.} Given a classical Hamiltonian function $H\in C^\infty(T^*Q)$, the \textit{KvH equation for classical wavefunctions} is (see \cite{BoGBTr} and references therein) 
\begin{equation}\label{KvH_eq}
{\rm i}\hbar\partial_t\Psi={\rm i}\hbar \{ H,\Psi\}- L_H  \Psi \qquad \text{with} \qquad L_H:={\mathcal{A}}\!\cdot\! X_H-H= p_i \partial_{p_i} H-H\,.
\end{equation}
Here, $X_H$ is the Hamiltonian vector field associated to $H$ so that $\mathbf{i}_{X_H}\omega={\rm d}H$, and $\{H,K\}=\omega(X_H, X_K)$ is the canonical Poisson bracket, extended to $\mathbb{C}$-valued functions by $\mathbb{C}$-linearity. In addition, we recognize that $L_H$ identifies the Lagrangian function (on phase-space) associated to the Hamiltonian $H$. 
The right hand side of \eqref{KvH_eq} defines the \emph{covariant Liouvillian operator}
\begin{equation}\label{preqop}
\widehat{\mathcal{L} }_H={\rm i}\hbar \{H,\ \}-L_H
\end{equation}
also known as \emph{prequantum operator}, which is easily seen to be an unbounded Hermitian operator on $\mathscr{H}_{\scriptscriptstyle C}$. As a consequence, the KvH equation \eqref{KvH_eq} is a Hamiltonian system with respect to the symplectic form \eqref{inner_symplectic} and the Hamiltonian
\begin{equation}\label{KvHham}
h(\Psi)= \int_{T^*Q}\bar\Psi \widehat{\mathcal{L} }_H\Psi\,{\rm d}z
\,.
\end{equation}

\smallskip\noindent
{\bf Lie algebraic structure.}
We remark that the correspondence $H\mapsto\widehat{\mathcal{L} }_H$ satisfies
\begin{equation}\label{LA_property}
[\widehat{\mathcal{L} }_H,\widehat{\mathcal{L} }_F]={\rm i}\hbar\widehat{\mathcal{L} }_{\{H,F\}}\,,
\end{equation}
for all $H,F\in C^\infty(T^*Q)$.
Hence, it follows that on its domain, the operator
\begin{equation}\label{LA_action}
\Psi\mapsto -{\rm i}{\hbar }^{-1}\widehat{\mathcal{L} }_H\Psi
\end{equation}
defines a  skew-Hermitian (or, equivalently, symplectic) left representation of the Lie algebra $(C^\infty(T^*Q),\{\ ,\ \})$ of Hamiltonian functions on $\mathscr{H}_{\scriptscriptstyle C}$.
We shall show that the Lie algebra action \eqref{LA_action} integrates into a unitary action of the group of strict contact transformations of \eqref{prequantumbundle} 
on $\mathscr{H}_{\scriptscriptstyle C}$, when the first cohomology group ${\sf H}^1(T^*Q, \mathbb{R})=0$ (or, equivalently, ${\sf H}^1(Q, \mathbb{R})=0$).

\smallskip\noindent
{\bf Madelung transform.} As discussed in the introduction, the KvH equation possesses an alternative formulation arising from the Madelung transform. By mimicking the quantum case, we write $\Psi=\sqrt{D} e^{{\rm i}S/\hbar}$ so that \eqref{KvH_eq} yields
\begin{equation}
\partial_t S+\{S,H\}= L_H
\,,\qquad
\partial_t D+\{D,H\}=0
\,.
\label{KvHMadelung2}
\end{equation}
Then, the first Madelung equation is revealing of the dynamics of the classical phase, which reads
\begin{equation}\label{classicalphase_evolution}
\frac{\de }{\de t}S(t,\eta(t,z))=L_H(\eta(t,z))\, , 
\end{equation}
where $\eta(t)$ is the flow of $X_H$.   
We remark that taking the differential of the first equation in \eqref{KvHMadelung2} leads to
\begin{equation}\label{dS_theta}
(\partial_t+\pounds_{X_H})({\rm d} S-{\mathcal{A}})=0\,,
\end{equation}
which is written in terms of the Lie derivative $\pounds_{X_H}={\rm d} \mathbf{i}_{X_H} + \mathbf{i}_{X_H}{\rm d}$.
At this point, one would be tempted to set $\de S={\cal A}$ so that the second equation in \eqref{KvHMadelung2} simply acquires the same meaning as the classical Liouville equation. However, things are not that easy: unless one allows for topological singularities in the phase variable $S$, the relation $\de S={\cal A}$ cannot hold and therefore the link between the KvH equation \eqref{KvH_eq} and the classical Liouville equation needs extra care. However, we shall come back to the Madelung picture later on to show how this may be extended beyond the current context.

\subsection{Strict contact transformations}
Having characterized the KvH equation and the Lie algebraic structure of prequantum operators, we now move on to characterizing their underlying group structure in terms of (strict) contact transformations and their unitary representation on Koopman wavefunctions.

\smallskip\noindent
{\bf Connection preserving automorphisms.} 
Given the trivial prequantum bundle \eqref{prequantumbundle}, we consider its automorphism group given by $\operatorname{Diff}(T^*Q)\,\circledS\, \mathcal{F}(T^*Q, S^1)$. There is a unitary representation of this group on $\mathscr{H}_{\scriptscriptstyle C}$ given by
\begin{equation}\label{sdp-action}
\Psi\mapsto U_{( \eta , e^{{\rm i} \varphi  })} \Psi = \sqrt{J_\eta^{-1}} ({e^{-{\rm i}\varphi/\hbar}}\Psi) \circ \eta ^{-1} \,,
\end{equation} 
where $(\eta,e^{{\rm i}\varphi})\in\operatorname{Diff}(T^*Q)\,\circledS\, \mathcal{F}(T^*Q, S^1)$. This is essentially the same representation as in \eqref{QuantumMadelungRep}, upon replacing the quantum configuration space $M$ with the classical phase space $T^*Q$.
A relevant subgroup of the automorphism group $\operatorname{Diff}(T^*Q)\,\circledS\, \mathcal{F}(T^*Q, S^1)$ is given by the group  of connection-preserving automorphisms of the principal bundle \eqref{prequantumbundle}: this group is given by
\begin{equation}\label{stricts}
\!\!\!\operatorname{Aut}_{\cal A}(T^*Q\times S^1)=
\big\{(\eta,e^{\rm i\varphi})\in\operatorname{Diff}(T^*Q)\,\circledS\, \mathcal{F}(T^*Q, S^1)\ \big|\ \eta^*\mathcal{A}+{\rm d}\varphi=\mathcal{A} \big\}
,
\end{equation}
where $\eta^*$ denotes pullback.
The above transformations were studied extensively  in van Hove's thesis \cite{VanHove} and are known as forming the group of \emph{strict contact diffeomorphisms}. Note that the relation $\eta^*\mathcal{A}+{\rm d}\varphi=\mathcal{A}$ implies
\begin{equation}\label{Aut-relations}
\eta^*({\rm d}{\cal A})=0
\,,\qquad\qquad
\varphi(z)=\theta+\int_{z_0}^{z}({\cal A}-\eta^*{\cal A})
\,,
\end{equation}
so that $\eta$ is a symplectic diffeomorphism, i.e. $\eta  \in \operatorname{Diff}_\omega(T^*Q)$. Also,  $\varphi$ is determined up to a constant phase $\theta=\varphi(z_0)$ since ${\sf H}^1(T^*Q, \mathbb{R})=0$ and thus the line integral above does not depend on the curve connecting  $z_0$ to $z$.
The Lie algebra of \eqref{stricts} is 
\begin{equation*}
\mathfrak{aut}_{\cal A}(T^*Q\times S^1)=
\big\{(X,{\nu})\in\mathfrak{X}(T^*Q)\,\circledS\, \mathcal{F}(T^*Q)\ \big|\ \pounds_X\mathcal{A}+{\rm d}\nu=0  \big\}
\end{equation*}
and we notice that this is isomorphic to the Lie algebra $ \mathcal{F} (T^*Q)$ endowed with the canonical Poisson structure. The Lie algebra isomorphism is
$
H \in \mathcal{F} (T^*Q) \longmapsto (X_H, -L_H ) \in  \mathfrak{aut}_{\cal A}(T^*Q\times S^1)
$.

\smallskip\noindent
{\bf The van Hove representation.} As a subgroup of the semidirect product $\operatorname{Diff}(T^*Q)\,\circledS\, \mathcal{F}(T^*Q, S^1)$, the group $\operatorname{Aut}_{\cal A}(T^*Q\times S^1)$ inherits from \eqref{sdp-action} a unitary representation, which is obtained by replacing \eqref{Aut-relations} in \eqref{sdp-action}. First appeared in van Hove's thesis \cite{VanHove}, we shall call this the \emph{van Hove representation}. Then, a direct computation shows that $-{\rm i}{\hbar }^{-1}\widehat{\cal L}_H$ emerges as the infinitesimal generator of this representation, i.e., we have
\begin{equation}\label{LG_action}
\left.\frac{d}{d\epsilon}\right|_{\epsilon=0} U_{(\eta_\epsilon, e^{{\rm i}\varphi  _\epsilon})}\Psi=-{\rm i}{\hbar}^{-1}\widehat{\cal L}_H  \Psi ,
\end{equation}
for a path $(\eta_\epsilon, e^{{\rm i}\varphi _\epsilon}) \in \operatorname{Aut}_{ \mathcal{A} }(T^*Q \times S ^1 )$ tangent to $(X_H, - L_H )$ at $(id, 1)$. Being an infinitesimal generator of the representation \eqref{sdp-action} restricted to $\operatorname{Aut}_{\cal A}(T^*Q\times S^1)$, $\widehat{\cal L}_H$ is equivariant, namely
\begin{equation}\label{equivariance_L_H}
U_{(\eta, e^{{\rm i} \varphi })}^\dagger \widehat{\cal L}_H U_{(\eta, e^{{\rm i} \varphi })}= \widehat{\cal L}_{H\circ\eta}\,, \quad \forall\;(\eta, e^{{\rm i} \varphi }) \in \operatorname{Aut}_{ \mathcal{A} }(T^*Q \times S ^1 )\,.
\end{equation}

\subsection{Momentum maps and the classical Liouville equation}

Having discussed the geometry underlying the KvH equation \eqref{KvH_eq}, we are now in the position of presenting its relation  to classical mechanics in terms of a momentum map taking Koopman wavefunctions to   distributions on phase-space \cite{BoGBTr}.
Since the representation \eqref{sdp-action} is unitary, it is symplectic with respect to the symplectic form \eqref{inner_symplectic} and admits a momentum map $ \mathcal{J} : \mathscr{H}_{\scriptscriptstyle C}\to\operatorname{Den}(T^*Q)$, where $\operatorname{Den}(T^*Q)$ denotes the space of density distributions  on $T^*Q$. From the general formula \eqref{momap_formula} for momentum maps for unitary representations, we have 
$\left<{\mathcal{J}}(\Psi),F\right>=\int_{T^*Q}\bar\Psi\,{\widehat{\cal L}_F \Psi}\,\de z$, 
%
where $\langle\ ,\,\rangle$ denotes the $L^2-$pairing between $\mathcal{F}(T^*Q)$ and its dual $\operatorname{Den}(T^*Q)$. This yields the expression
\begin{align}
\mathcal{J} (\Psi)=&\ |\Psi|^2 - \operatorname{div}\!\big( \mathbb{J} {\mathcal{A}} |\Psi|^2\big)  + {\rm i} \hbar\{\Psi,\bar\Psi\}
\nonumber
\\
=&\ |\Psi|^2 - \operatorname{div}\!\big(\bar \Psi \mathbb{J}( {\mathcal{A}} \Psi + {\rm i} \hbar\nabla\Psi)\big)\,.
\label{KvHmomap}
\end{align} 
Here, the divergence is associated to the Liouville form and $\mathbb{J}:T^*(T^*Q)\rightarrow T(T^*Q)$ is defined by $\{F,H\}= \langle {\rm d}F, \mathbb{J}({\rm d}H)\rangle$.
This equivariant momentum map is a Poisson map with respect to the symplectic Poisson structure $\{\!\!\{f,h\}\!\!\}(\Psi)$ on $\mathscr{H}_{\scriptscriptstyle C}$ and the Lie-Poisson structure $\{\!\!\{f,h\}\!\!\}(\rho)$ on $\operatorname{Den}(T^*Q)$. With an abuse of notation, we write
\[
\{\!\!\{f,h\}\!\!\}(\Psi)=\frac1{2\hbar}\operatorname{Im}\int_{ T^*Q} \overline{\!\frac{\delta f}{\delta \psi}} \, \frac{\delta h}{\delta \psi}\,{\rm d}z 
\quad\longrightarrow\quad
 \{\!\!\{f,h\}\!\!\}(\rho)=\int_{T^*Q}\!\!
\rho\left\{\frac{\delta f}{\delta \rho},\frac{\delta h}{\delta \psi}\right\} {\rm d}z\,.
 \] 
Hence, if $\Psi(t)$ is a solution of 
the KvH equation, the density $ \rho(t)  = \mathcal{J} (\Psi (t))$ in \eqref{KvHmomap} solves the Liouville equation
$\partial_t\rho=\{H,\rho\}$.
A density of the form \eqref{KvHmomap} is not necessarily positive definite. However, the Liouville equation generates the sign-preserving evolution $\rho(t)=\eta(t)_*\rho_0$, where $\eta(t)$ is the flow of $X_H$, thereby recovering the usual probabilistic interpretation. In summary, we have the following scheme:
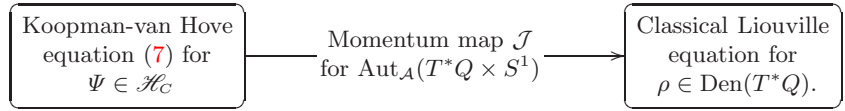
\vspace{-0.3cm}\begin{figure}
{\small\hspace{-0.5cm}
\begin{xy}
\xymatrix{
&*+[F-:<3pt>]{\begin{array}{c}\text{Koopman-van Hove}\\
\text{equation \eqref{KvH_eq} for}\\
\text{ $\Psi \in \mathscr{H}_{\scriptscriptstyle C}$}
\end{array}}\ar[rrrrr]|{\begin{array}{c}
\text{Momentum map $\mathcal{J}$}\\
\text{for $\operatorname{Aut}_{\cal A}(T^*Q \times S ^1 )$} 
\end{array}
} & & & &&
*+[F-:<3pt>]{
\begin{array}{c}
\text{Classical Liouville}\\
\text{equation 
for}\\
\text{ $ \rho  \in \operatorname{Den}(T^*Q)$.}
\end{array}
}
}
\end{xy}
}
\vspace{-0.3cm}
\end{figure}

\noindent
Having characterized the classical phase-space density in terms of Koopman wavefunctions \eqref{KvHmomap}, we ask if this representation is sufficient to reproduce all the possible classical densities. For example, Gaussian densities allow for this representation \cite{BoGBTr}. However, this may not be true in more general cases for which KvH theory needs to be extended appropriately.



\section{Hydrodynamic quantities and von Neumann operators \label{sec:KvHMadelung}} 
Let us start our discussion by  introducing Madelung's hydrodynamic quantities
\[
\big(\hbar\operatorname{Im}(\Psi^*\nabla\Psi),|\Psi|^2\big)=(D\nabla S,D)=:(\sigma,D)
\]
as they arise from applying the momentum map \eqref{Madmomap} to the Koopman wavefunction. We notice that the KvH Hamiltonian \eqref{KvHham} is rewritten in terms of these variables as
\begin{equation}\label{MadKvHHam}
h(\sigma,D)=\int_{T^*Q}(X_H\cdot\sigma-DL_H)\,\de z
\,.
\end{equation}
Then, upon using the results in Section \ref{sec:KvH}, we write the evolution equations as
\begin{equation}
(\partial_t+\pounds_{X_H})(\sigma-D{\mathcal{A}})=0\,,
\qquad\quad
\partial_tD+\operatorname{div}(D X_H)=0
\,.
\label{LPDynKvH}
\end{equation}
These variables provide an alternative representation of the classical  density via the momentum map \eqref{KvHmomap}, that is
\begin{equation}
\rho=D +\operatorname{div}( \mathbb{J}\sigma-\mathbb{J}{\mathcal{A}} D)
\,.
\label{LiouvillemomapVN}
\end{equation}
Once again, one is tempted to set $\sigma=D{\cal A}$ so that $\rho=D$, thereby eliminating all possible restrictions arising from specific representations of the classical density. However, we have $\de(\sigma/D)=0$ and even if one allows for topological singularities in the phase so that $\nabla S={\cal A}$ and $\rho=|\Psi|^2$, the last relation prevents the existence of $\delta-$like particle solutions. These observations lead us to the necessity of extending the present construction in order to include more general classical distributions.

\smallskip\noindent
{\bf Von Neumann operator.} Here, we exploit the analogy between classical and quantum wavefunctions to introduce a self-adjoint von Neumann operator  $\widehat{\Theta}$. We write the von Neumann evolution equation \cite{BoGBTr}
\begin{equation}\label{VNeq}
{\rm i}\hbar\frac{\partial\widehat{\Theta}}{\partial t}=\big[\widehat{\cal L}_{H},\widehat{\Theta}\,\big],
\end{equation}
so that $\widehat{\Theta}=\Psi\Psi^\dagger$ recovers that KvH equation \eqref{KvH_eq}. Here, the adjoint is defined so that $\Psi_1^\dagger\Psi_2=\langle\Psi_1|\Psi_2\rangle$. However, in the present discussion, we allow for a general von Neumann operator $\widehat{\Theta}$ with unit trace, while we do not necessarily restrict $\widehat{\Theta}$ to be positive-definite, although this can be set as a convenient initial condition. As is well known, equation \eqref{VNeq} is a Hamiltonian system with the following  Lie-Poisson structure:
\begin{equation}\label{VNLP}
\{\!\!\{f,h\}\!\!\}(\widehat{\Theta})=-{\rm i}\hbar^{-1}\operatorname{Tr}\left(\widehat{\Theta}\left[\frac{\delta f}{\delta \widehat{\Theta}},\frac{\delta h}{\delta \widehat{\Theta}}\right]\right)
\,,\qquad\qquad
h(\widehat{\Theta})=\operatorname{Tr}\big(\widehat{\cal L}_{H}\widehat{\Theta}\big)
\,.
\end{equation}

\smallskip\noindent
{\bf Momentum map.} 
As shown in \cite{FoHoTr}, the representation \eqref{QuantumMadelungRep} on the wavefunction transfers naturally to a unitary action on the von Neumann operator that is conveniently written in terms of its integral kernel ${\cal K}_{\widehat{\Theta}}(z,z')$. Indeed, one writes
\[
{\cal K}_{\widehat{\Theta}}(z,z')\mapsto \sqrt{J_\chi^{-1}(z)J_\chi^{-1}(z')}\,{e^{-{\rm i}\hbar^{-1}(\varphi(z)-\varphi(z'))}}{\cal K}_{\widehat{\Theta}}(\chi^{-1}(z),\chi^{-1}(z'))
\,,
\]
where $( \chi  , e^{{\rm i} \varphi} ) \in \operatorname{Diff}(T^*Q) \,\circledS\, \mathcal{F} (T^*Q, S ^1 )$. In turn, this action produces the equivariant momentum map
\begin{align}\nonumber
&\mathcal{J} ( \widehat\Theta )(z)=\Big(\frac{{\rm i}\hbar}2\frac{\partial}{\partial z}{\cal K}_{\widehat{\Theta}}(z',z)-\frac{{\rm i}\hbar}2\frac{\partial}{\partial z}{\cal K}_{\widehat{\Theta}}(z,z'),\,{\cal K}_{\widehat{\Theta}}(z,z')\Big)\Big|_{z'=z}=(\sigma(z),D(z))
\label{Madmomap2}
\end{align}
and we verify that the Hamiltonian $h(\widehat{\Theta})$ in \eqref{VNLP} may  be rewritten exactly as in \eqref{MadKvHHam}. Then, as in the case of quantum hydrodynamics, $\mathcal{J}(\widehat{\Theta})$ is a Poisson map hence mapping the von Neumann equation \eqref{VNeq} to the Lie-Poisson dynamics \eqref{LPDynKvH} on  $\mathfrak{X} (T^*Q)^* \times \operatorname{Den} (T^*Q)$. 

\smallskip\noindent
{\bf Point particles in phase-space.} 
At this point, we have $\de(\sigma/D)\neq0$ and the relation $\sigma=D{\cal A}$ can be adequately set as an initial condition that is preserved in time. In this case, the momentum map \eqref{LiouvillemomapVN} representing the classical density reduces to $\rho=D$, which now is allowed to have the $\delta-$like expression  $D(z)=\delta(z-\zeta)$
reproducing a point particle located at the phase-space coordinates $\zeta$.  Then, one may wish to identify a possible solution of the von Neumann equation \eqref{VNeq} ensuring that the relation $\sigma=D{\cal A}$ is indeed satisfied. Following the discussion in \cite{FoHoTr}, one can see that this is given by the integral kernel
\begin{align*}
{\cal K}_{\widehat{\Theta}}(z,z')\!=\!&\ D\Big(\frac{z+z'}2\Big)\exp\!\Big[\frac{\rm i}\hbar(z-z')\!\cdot\!{\cal A}\Big(\frac{z+z'}2\Big)\Big]\!=\! D\Big(\frac{z+z'}2\Big)\,e^{\frac{\rm i}{2\hbar}(p+p')\cdot(q-q')},
\end{align*}
where the second step uses ${\cal A}=p_k\de q^k$. The origin of this (unsigned) form of the von Neumann operator may be unfolded by resorting to the Wigner function formalism, although this discussion would take us beyond the purpose of the present work.

\end{document}